\documentstyle[twocolumn,aps,epsf]{revtex}
\newcommand{\beq}{\begin{equation}}
\newcommand{\eeq}{\end{equation}}
\newcommand{\beqa}{\begin{eqnarray}}
\newcommand{\eeqa}{\end{eqnarray}}

\newcommand{\opa}{\hat{A}}
\newcommand{\opb}{\hat{B}}
\newcommand{\opd}{\hat{D}}

\begin{document}
\draft
\title{Simultaneous minimum-uncertainty measurement of discrete-valued complementary observables}

\author{Alexei Trifonov,\footnote{Electronic address: alexei@ele.kth.se\\
Permanent address: Ioffe Physical Technical Institute, 26
Polytechnicheskaya,194021 St. Petersburg, Russia.} Gunnar Bj\"{o}rk, and Jonas
S\"{o}derholm}
\address{Department of Electronics, Royal Institute of Technology (KTH), Electrum 229, SE-164 40 Kista, Sweden}
\date{\today}
\maketitle

\begin{abstract}
We have made the first experimental demonstration of the simultaneous minimum uncertainty product between two complementary observables for a two-state system (a qubit). A partially entangled two-photon state was used to perform such measurements. Each of the photons carries (partial) information of the initial state thus leaving a room for measurements of two complementary observables on every member in an ensemble.
\end{abstract}


%03.65.Bz = theory of measurement in quantum mechanics
%42.50.-p = Quantum optics
\narrowtext

Complementarity and the associated uncertainty relations play a key role in quantum theory. Uncertainty can be quantified in several ways, e.g., in terms of entropy \cite{Deutsch}, variance \cite{Schrodinger,Robertson,Puri,Bjork 2}, or other quadratic functions of measurement probabilities \cite{Englert,Zeilinger}. However, what is usually meant by "uncertainty" is the state's {\em inherent} indeterminism with respect to any two complementary observables \cite{Zeilinger,Pauli}. A lower bound for this inherent indeterminism is given by the ordinary Schr\"{o}dinger-Robertson uncertainty relation \cite{Schrodinger,Robertson} which is based on the variances obtained from a sharp measurement of one of the observables on a particular ensemble, and a sharp measurement of the complementary observable on {\em another}, but identically prepared ensemble. For example, to determine the variance of one quadrature amplitude of a electromagnetic field, the field can be mixed with a local oscillator field with an appropriately adjusted relative phase in a 50/50 beam splitter. A signal proportional to the quadrature amplitude is then obtained by putting photo-detectors at the two beam splitter output ports and subtracting the respective photo-currents \cite{Yuen}. To obtain the variance of the complementary quadrature component the experiment is repeated on {\em another}, identically prepared, field, but with the local oscillator relative phase shifted by $\pi/2$. Experimentally the Schr\"{o}dinger-Robertson uncertainty relation has been tested in exactly this way \cite{Kimble}.

However, if one wishes to determine the values of two complementary operators on every single member of an ensemble of identically prepared particles, e.g., the value of the spin in both the $x$ and the $y$ directions of a single particle, then additional complications arise. Should one measure the spin $x$ component by making a sharp measurement, then a subsequent measurement of the $y$ component will yield a completely random outcome, irrespective of how the particle was prepared initially. Consequently, if we want some correlation between the outcome of the spin $y$ measurement and the preparation of the particle, the measurement of the spin $x$ component must not be sharp.

The first to address this question were Arthurs and Kelly \cite{Arthurs I,Arthurs II} who studied canonical observables, such as position and momentum. The objective was to find a minimum uncertainty relation for the observables based on the measurement outcomes of {\em both} observables on {\em every} member of the ensemble. However, in a simultaneous measurement one has two conflicting objectives: The measurement variances should be minimized while the correlations between the measurement outcomes and the preparation of the ensemble should be maximized. There is no unique way of doing this. Arthurs and Kelly added an additional requirement, namely that the expectation values of the two (unsharp) measurements should equal the  expectation values of the corresponding {\em sharp} measurements on any ensemble of identically prepared particles. With this additional requirement Arthurs and Kelly found an uncertainty relation for canonical observables \cite{Arthurs I} equal in form to the Schr\"{o}dinger-Robertson relation, but where the minimum product of the variances was four times greater than the Schr\"{o}dinger-Robertson bound. They also found that in order to achieve this lower bound, {\em a priori} knowledge about the preparation of the ensemble was needed to adjust the relative sharpnesses of the two measurements. (The sharper one measurement is made, the more unsharp the other must become). To the best of our knowledge, Arthurs and Kelly's relation has never been tested experimentally.

\begin{figure}[h]
\leavevmode
\epsfxsize=8cm
\epsfbox{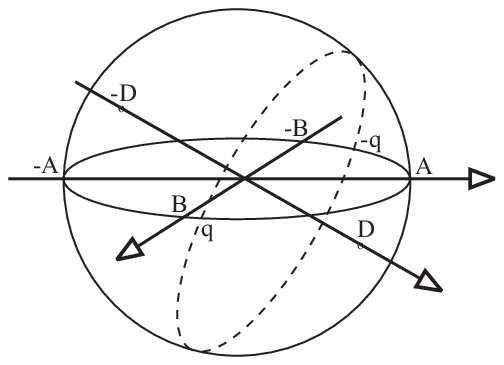}
\vspace{0.3cm}
\caption{The Bloch sphere associated with $|\psi\rangle$, $\opa$, and $\opb$.}
\label{Fig:Bloch}
\end{figure}

In an earlier work we have built on Arthurs and Kelly's work and derived a similar simultaneous uncertainty relation for two-state systems, i.e., for two non-canonical observables \cite{Bjork 2}. Any pure state in the associated Hilbert space can be represented as a point on the Bloch sphere. Suppose we want to measure two complementary observables $\opa$ and $\opb$ and that the Bloch sphere is oriented so that $\opa$'s and $\opb$'s respective eigenstates $|A_+ \rangle$, $|A_- \rangle$, and $|B_+ \rangle =(|A_+ \rangle + |A_- \rangle)/\sqrt{2}$, and $|B_- \rangle =(|A_+ \rangle - |A_- \rangle)/\sqrt{2}$ lie equidistantly around the equator, see Fig. \ref{Fig:Bloch}. Also suppose that the corresponding eigenvalues are $\pm A$ and $\pm B$. It is easy to show that of all the states lying on any specific ``longitude'' on the Bloch sphere, the state defined by the point where the longitude crosses the equator will have the smallest uncertainty product between $\opa$ and $\opb$. Hence, we will restrict our attention to the equatorial states that all have the general form

\beq
\left| \psi  \right\rangle  = \sqrt {w_{A+}} \left| {A_ +  } \right\rangle  \pm \sqrt {w_{A-}} \left| {A_ -  } \right\rangle .
\label{eq:initial state}
\eeq
The positive, real numbers $w_{A + } $  and $w_{A - }=1- w_{A + }$  are the probabilities of obtaining the results $A$ and $-A$ if we make a sharp $\opa$ measurement on the state. 

Hermitian operators with discrete finite spectra cannot be canonical, i.e., have a commutator of the form $i C$, where $C$ is a real number. Therefore, their minimum uncertainty product is state dependent and not fixed \cite{Schrodinger,Robertson,Puri}. For example, the minimum product of the sharply measured variances vanishes for the eigenstates of $\opa$ and $\opb$ since one of the variances is zero while the other is finite. However, for {\em any} state on the equator (that is, for any value $0 \leq w_{A+} \leq 1$), there exists a combination of simultaneous unsharp $\opa$ and $\opb$ measurements that reaches the minimum variance product dictated by the appropriate simultaneous uncertainty relation \cite{Puri,Bjork 2}. It is our objective in this paper to find and make these measurements.

Let us first show that there is no von Neumann measurement that will give us simultaneous knowledge about $\opa$ and $\opb$ and still fulfill our ``correct expectation value'' requirement. Consider such a measurement of an observable $\opd \neq \opa$, $\opb$ in the space. The measurement is characterized by an axis through the origin of the Bloch sphere, crossing the sphere shell at points $D$ and $-D$. However, all states in a plane normal to the $D$-axis will give the same measurement probability distributions. In particular, this holds for the two states defined by the crossing points $q$ and $-q$ between the plane and the equator in spite of them having different values of $\langle \opa \rangle$, $\langle \opb \rangle$, or both. This makes it impossible to fulfill the ``correct expectation value'' requirement, so we must rule out any von Neumann measurements and instead look at positive operator valued measurements, POMs. 

To implement a POM, we need to enlarge our state space. A sufficient way in our case is to introduce an ancillary two-state particle, which we henceforth will call the probe (the particle whose characteristics we wish to measure will be called the object). In order to extract any information about the object from the probe, we must entangle the two particles. We shall assume that the (unitary) entanglement interaction affects only the probe's state $| m \rangle$, v.i.z.
\begin{eqnarray}
\hat{U} | A_{+} \rangle \otimes | m \rangle  &=& 
|A_{+} \rangle \otimes | m_{+}\rangle ,  \label{eq:unitary_transf} \\
\hat{U} | A_{-} \rangle \otimes | m \rangle  &=& 
|A_{-} \rangle \otimes | m_{-}\rangle ,  \nonumber
\end{eqnarray}
which, for the initial state $|\psi\rangle$, results in the state
\begin{equation}
|\psi _{e}\rangle =\sqrt{w_{A+}}|A_{+}\rangle \otimes |m_{+}\rangle \pm \sqrt{w_{A-}}|A_{-
}\rangle \otimes |m_{-}\rangle .  \label{eq:entangled system}
\end{equation}
A measure of the entanglement of $|\psi _{e}\rangle$ is $c=|\langle m_{+}|m_{-}\rangle |$ where $c=0$ ($c=1$) signifies perfect (no) entanglement.

Suppose we decide to infer knowledge about $\hat B$  from a direct measurement on the object particle, and infer information about $\hat A$ (of the object) from the probe particle.  Three problems arise: how to ensure that the inferred (and therefore unsharply) measured mean values equal the true (sharp) means, how to chose basis to measure the probe state, and how to optimally choose the entanglement parameter $c$?

All questions were addressed and answered in \cite{Bjork 2}. Here we will only recapitulate the main conclusions. A sharp measurement of $\opb$ on the object particle in state (\ref{eq:entangled system}) will yield either of two outcomes $B$ or $-B$ with probabilities $w_{B+}'$ and $w_{B-}'$ that will in general not equal $w_{B+}\equiv|\langle \psi|B_+ \rangle|^2=1/2 \pm \sqrt{w_{A+}w_{A-}}$ and $w_{B-}\equiv|\langle \psi|B_- \rangle|^2=1-w_{B+}$. However, if we associate the measurement outcomes with the rescaled eigenvalues $\pm B/c$, then the expectation values of a sharp  rescaled $\opb$ measurement on object particle in state $|\psi_e\rangle$ and a sharp $\opb$  measurement on $|\psi\rangle$ become equal [for any state of the form (\ref{eq:entangled system})]. The former measurement is an inferred and unsharp estimation of the latter.

Let us now consider the indirect measurement of $\hat{A}$. Projecting the probe state on either of the projectors $| m_{+}\rangle \langle m_{+}|$ or $| m_{-}\rangle \langle m_{-}|$, we can get the probabilities $w_{A+}$ and $w_{A-}$. However, it is not possible to obtain both $w_{A+}$ and $w_{A-}$ exactly by any fixed measurement on the probe state because of the non-orthogonality of $\left| m_{+}\right\rangle $ and $\left|
m_{-}\right\rangle $ (if $c \neq 0$). The probabilities will be optimally estimated by using an orthogonal basis derived in \cite{Bjork 2}. It can be shown that the corresponding von Neumann measurement basis
vectors $\left| M_{+}\right\rangle$ and $\left|
M_{-}\right\rangle $ then form equal angles $\gamma$ with the vectors $\left|
m_{+}\right\rangle $ and $\left| m_{-}\right\rangle $ respectively (see Fig. \ref{Fig:probe basis}). The probabilities $w_{A+}'$ and $w_{A-}'$ of obtaining the results $M_{+}$ and $M_{-}$ are not equal to $w_{A+}$ and $w_{A-}$ (unless $c=0$). However, if we associate the respective probe particle measurement outcomes with the rescaled eigenvalues $ \pm A/\sqrt{1-c^{2}}$ the inferred (unsharp) measurement of $\opa$ through $|\psi_e\rangle$ will give the true mean of $\opa$ measured on {\em any} state $|\psi\rangle$ irrespective of the value of $w_{A+}$.

Now, how do we obtain a minimum uncertainty product for a simultaneous $\hat{A}$ and $\hat{B}$ 
measurement? For a given state $|\psi\rangle$ the only adjustable parameter in $|\psi_e\rangle$ is the degree of entanglement $c$. Thus the simultaneous uncertainty product must be minimized with respect to $c$ for a given $w_{A+}$, i.e., $|\psi\rangle$ must be known {\em a priori}.

Let $\delta \hat{A}=(\langle  \Delta \hat{A}^{2} \rangle)^{1/2}/A$ and $\delta \hat{B}=(\langle  
\Delta \hat{B}^{2} \rangle)^{1/2}/B$ be the normalized uncertainties of $\hat{A}$ and 
$\hat{B}$ when $|\psi\rangle$ is measured. The normalized uncertainties, inferred from simultaneous measurements on $|\psi_e\rangle$,  are denoted $\delta 
\hat{A}^{\prime }$ and $\delta \hat{B}^{\prime }$. A straightforward calculation \cite{Bjork 2} gives
\begin{equation}
\delta \hat{A}^{\prime }=\sqrt{(\delta \hat{A})^{2}+ c^{2}/(1-c^{2})} ,
\label{eq:uncertainty_meas_A} 
\end{equation}
and
\begin{equation}
\delta \hat{B}^{\prime }=\sqrt{(\delta \hat{B})^{2}+ (1-c^{2})/c^{2}} .
\label{eq:uncertainty_meas_B}
\end{equation}
The product of (\ref{eq:uncertainty_meas_A}) and (\ref{eq:uncertainty_meas_B}) reaches the 
minimum
\begin{equation}
(\delta \hat{A}^{\prime }\delta \hat{B}^{\prime })_{\min }=1+\delta \hat{A}\delta \hat{B}, \mbox{ for } c=\sqrt{\delta \hat{A}/(\delta \hat{A}+\delta \hat{B})} .
\label{eq:UP product min}
\end{equation}
These formulae are identical to (53) and (54) in \cite{Bjork 2} but are expressed in different 
quantities. We see that the minimum product of the simultaneously measured observables' variances is {\em at least} nine times larger than the corresponding product of the sharply measured observables (since $0\leq\delta \opa\delta \opb\leq 1/2$).

Now we would like to test the simultaneous uncertainty relation with photon states. Unfortunately, state-of-the-art technology does not allow us to perform the unitary interaction (\ref{eq:unitary_transf}) on single photons. Therefore, we start directly with the state (\ref{eq:entangled system}) without introducing independent object and probe states first. This is not such a serious flaw as it may appear. If we had started with $|\psi\rangle$, then we would have had to make sure that the state after the entanglement was indeed $|\psi_e\rangle$, by measuring, e.g., $w_{A+}$, $c$ and the relative phase-angle between the state's two terms (which, if the entangling step works alright, is akin to measuring $w_{B+}$). Hence, there is actually little point in using a two-step procedure to arrive at $|\psi_e\rangle$ since the parameters $w_{A+}$ and $w_{B+}$ (uniquely defining $|\psi\rangle$) can, and must, be measured from $|\psi_e\rangle$ anyhow. (In a related work \cite{Trifonov 2} we describe how we could measure $w_{B+}$ by a so called quantum erasure measurement.)

Experimentally, the state (\ref{eq:entangled system}) was created via non-collinear spontaneous 
parametric down-conversion  with type II phase matching. Our source was similar to that described 
in \cite{Kwiat}, but we used a pulsed source to reduce the random coincidence count-rate. With a specific choice of the relative phase between the vertically $|\uparrow\rangle$ and horizontally $|\rightarrow\rangle$ polarized photons the state after the crystal is ideally in a mixture of the vacuum state, one photon states (that both are eliminated by post-selection through detector correlation) and the state
\begin{equation}
| \psi_i \rangle = \left ( | \uparrow
\rangle \otimes | \rightarrow \rangle - | \rightarrow \rangle \otimes |\uparrow
\rangle \right )/ \sqrt{2}.
\end{equation}
If we choose the leftmost ket in the product to represent the object and the second to represent the probe, the state becomes exactly the state (\ref{eq:entangled system}) with $w_{A+}=w_{A-}=\frac{1}{2}$ and $c=0$. To adjust the state for a minimum simultaneous uncertainty product measurement, a partial polarizer was inserted in the object spatial mode, rotated at an angle $\alpha $ with respect to the horizontal plane. The partial polarizer consisted of a stack of $N$ glass plates held at the Brewster angle with respect to the object photon propagation axis. The amplitude transmittivities of the partial polarizer were $t_{{\rm p}}\approx 1$ and $t_{{\rm s}}=t$, where $t$ was determined by the number of plates $N$. (The indices $\rm p$ and $\rm s$ refer to the linear polarization states with the respect of the partial polarizer.) Assume that we insert the partial polarizer so that it tends to polarize the object photon vertically (corresponding to $|A_+\rangle$). In this case $c$ will remain zero, but $w_{A+}$ of the {\em post selected} states will become larger than $w_{A-}$. If, on the other hand, we make the polarizer perfect and insert it at $\alpha=\pi/4$ then $c$ will become unity but $w_{A+}$ and $w_{A-}$ remain equal. Hence, the degree of entanglement between the object and probe, as well as the {\em a priori} path probabilities could be varied by changing the two degrees of freedom, $N$ and $\alpha $, of the partial polarizer. Unfortunately, with this device, it is not possible to vary $c$ and $w_{A+}$ independently. That is, for a given number of plates, only two choices of $\alpha $ give the combination of $c$ and $w_{A+}$ that makes the simultaneous uncertainty product attain its lowest bound.

The experimental setup is shown in Fig. \ref{Fig:experimental setup}. A 0.5 mm long beta-barium borate (BBO) crystal was used to generate non-colinear frequency degenerate photon pairs in a spontaneous down conversion process. The pump was 100 mW average power second harmonic generation from the initial 2 W average power Ti-Sapphire pulsed radiation. The experimental parameters were 2 ps pulse duration, 80 MHz repetition rate, and 390 nm wavelength of the second harmonic. The state detection was accomplished by two EG\&G single photon detectors (SPC1 and SPC2) with around 60\% quantum efficiency and a coincidence counter. The photon pair was selected by two 1 mm diameter irises placed 1 m from the crystal, selecting pairs from a 5 degree top angle emission cone. Identical 10 nm bandpass filters (F1 and F2) centered at $\lambda=780$ nm were placed in front the detectors to select frequency degenerate photon pairs. This resulted in 10 kHz single counts events and around 100 Hz of coincidence counts. Two polarizers (P1 and P2) were used to select linearly polarized photons. Polarization rotation of the beams was accomplished by two $\lambda /2$ plates placed before the polarizers. Because the uncertainty product is symmetrical in $w_{A+}$ with respect to the $w_{A+}=0.5 $ point, only the $0.5 \leq w_{A+} \leq 1$ probability range was used for the minimum uncertainty product measurements. Three different numbers of 
glass plates ($N=7$, 8, and 10) were used to adjust the state $| \psi_i \rangle$. This choice gave more or less an equally spaced set of probabilities $w_{A+}$. Thus six combinations of $N$ and $\alpha $ should give the minimum uncertainty product. For a given number of glass plates held at the theoretically computed value of $\alpha$ the four coincidence probabilities with respect to the probe measurement basis orientation were measured, from which the values of $c$ and $w_{A+}$ at this particular setting were derived. This calibration curve allowed us to adjust the partial polarizer rotation angle $\alpha $ and thus the values of $c$ and $w_{A+}$ rather precisely in order to attain the minimum uncertainty product bound.

To measure the $\hat{B}$ outcome probabilities the photon counter, a $\lambda /2$ 
plate, and a polarizer, were used to project the object state onto the $\left| B_{+}\right\rangle $ 
and $\left| B_{-}\right\rangle $ eigenstates. (These are the states polarized at 45 and 135 degrees from the horizontal direction.) The $\delta\hat{B}'$ uncertainty was calculated using the 
measured probabilities with subsequent rescaling due to the measured entanglement parameter $c$. The $\delta\hat{A}'$ uncertainty was measured by orienting 
the probe measurement basis to coincide with $|M_+\rangle$ and $|M_-\rangle$ (see Fig. \ref {Fig:probe basis}) which are also governed by the entanglement parameter $c$. After subsequent rescaling, the uncertainty product was calculated and plotted in Fig. \ref{Fig:results}, which is the main result of this paper. The six circles represent the measured uncertainty product. The result is in relatively good agreement with the theoretical estimation (\ref{eq:UP product min}) shown by the solid line.
The slight increase of the experimentally obtained values with respect to the theoretical prediction was mainly due to the imperfect purity of the 
prepared state (\ref{eq:entangled system}). The degradation was mainly due to the group 
dispersion effects in the BBO crystals and light scattering from the numerous optical 
elements. The maximum uncertainty product
was calculated using the assumption that both the $\hat{A}$ and $\hat{B}$ measurement's two respective 
outcomes had equal probabilities (that is, as if the initial object state lie at the north or south pole of the Bloch sphere), and using the same rescaling as for the minimum uncertainty 
product calculation. Then the maximum uncertainty product is given by 
\begin{equation}
(\delta \hat{A}^{\prime }\delta \hat{B}^{\prime})_{\max }=(c \sqrt{1-c^{2}})^{-1},
\end{equation}
and is shown as a reference by the dashed line.

In conclusion we have experimentally verified, to the best of our knowledge for the first time, a simultaneous uncertainty relation, namely that between discrete non-canonical observables. This result has a general significance in the context of measurement theory and complementarity, and it underlines the specific behavior of non-canonical discrete operators. 

This work was financially supported by INTAS and by the Royal Swedish Academy of Sciences (KVA).

\section{Figure Captions}

\begin{figure}[h]
\caption{The probe measurement basis.}
\label{Fig:probe basis}
\end{figure}

\begin{figure}[h]
\caption{A schematic illustration of the experimental setup.}
\label{Fig:experimental setup}
\end{figure}

\begin{figure}[h]
\caption{Results of a simultaneous minimum uncertainty product measurement (circles). 
The theoretical prediction (\ref{eq:UP product min}) is shown by the solid line. 
The maximum uncertainty product (dashed line) and the Schr\"{o}dinger-Robertson uncertainty relation (dotted line) are given as references.}
\label{Fig:results}
\end{figure}


\bigskip

\begin{references}

\bibitem{Deutsch} D. Deutsch, \prl {\bf 50}, 631 (1983); H. Maassen and J. B. M. Uffink, \prl  {\bf 60}, 1103 (1988); J. S\'{a}nchez, Phys. Lett. A  {\bf 173}, 233 (1992); M. J. W. Hall, \prl {\bf 74}, 3307 (1995).

\bibitem{Schrodinger}  E. Schr\"{o}dinger, Proc. Prussian Academy of Sciences Physics-Mathematical Section, {\bf XIX}, 296 (1930). An English
translation can be found in Los Alamos e-print archive quant-ph/9903100.

\bibitem{Robertson}  H. P. Robertson, Phys. Rev. {\bf 35}, 667A (1930); {\bf 46}, 794 (1934).

\bibitem{Puri}  R. R. Puri, Phys. Rev. A {\bf 49}, 2178 (1994).

\bibitem{Bjork 2}  G. Bj\"{o}rk {\em et al.}, \pra {\bf 60}, 1874 (1999).

\bibitem{Englert} B.-G. Englert, \prl {\bf 77}, 2154 (1996).

\bibitem{Zeilinger}  \v{C}. Brukner and A. Zeilinger, \prl {\bf 83}, 3354
(1999); Los Alamos e-print archive quant-ph/0006087; M. J. W. Hall, {\em ibid} quant-ph/0007116.

\bibitem{Pauli} W. Pauli, Die allgemeinen Prinzipien der Wellenmechanik in {\em Handbuch der Physik} Band V, 1, Hrsg. S. Fl\"{u}gge, Springer-Verlag, Berlin 1990.

\bibitem{Yuen}  H. P. Yuen and J. H. Shapiro, IEEE Trans. Inf. Theory {\bf 26}, 78 (1980).

\bibitem{Kimble}  L.-A. Wu, M. Xiao, and H. J. Kimble, J. Opt. Soc. Am. B {\bf 4}, 1465 (1987).

\bibitem{Arthurs I}  E. Arthurs and J. L. Kelly Jr., Bell Syst. Tech. J. {\bf 44}, 725 (1965).

\bibitem{Arthurs II}  E. Arthurs and M. S. Goodman, \prl
{\bf 60}, 2447 (1988).

% \bibitem{Zeilinger 2} A. Zeilinger, Reviews of Modern Physics,  {\bf 71},  (1999)

% \bibitem{Trifonov}  D. A. Trifonov, J. Math. Phys. {\bf 35}, 2297 (1994).

\bibitem{Trifonov 2}  A. Trifonov, G. Bj\"{o}rk, J. S\"{o}derholm, and
T. Tsegaye, Los Alamos e-print archive quant-ph/0009097.

\bibitem{Kwiat} P. Kwiat {\em et al.}, \prl {\bf 75}, 4337 (1995).


\end{references}
\end{document}